# Domain wall motions in perpendicularly magnetized CoFe/Pd multilayer nanowire

Zhaoliang Meng,[1,2] Manoj Kumar,[1] Jinjun Qiu,[2] Guchang Han,[2,*] Kie Leong Teo[1,2] and Duc-The Ngo,[3,1,*]

[1]Department of Electrical and Computer Engineering, National University of Singapore, 4 Engineering Drive 3, Singapore 117576

[2]Data Storage Institute, A*STAR (Agency for Science Technology and Research), 5 Engineering Drive 1, Singapore 117583

[3]Department of Micro- and Nanotechnology, Technical University of Denmark, Kgs. Lyngby 2800, Denmark

**Abstract**

Current induced domain wall (DW) motion has been investigated in a 600-nm wide nanowire using multilayer film with a structure of Ta(5 nm)/Pd(5 nm)/[CoFe(0.4 nm)/Pd(1.2 nm)]$_{15}$/Ta(5 nm) in terms of anomalous Hall effect measurements. It is found that motion of DWs can be driven by a current density as low as $1.44 \times 10^{11}$ Am$^{-2}$. The effects of the Oersted field ($H_{Oe}$) and spin transfer torque field ($H_{ST}$) on the wall motion, which are considered as effective fields, have been quantitatively separated from the dependence of depinning fields on the current. The results show that the motion of the walls was essentially dominated by the non-adiabaticity with a high non-adiabatic factor $\beta$ up to 0.4.

[*] Corresponding authors: **Dr. Han Guchang** (HAN_Guchang@dsi.a-star.edu.sg) and **Dr. Duc-The Ngo** (dngo@nanotech.dtu.dk)

## I. INTRODUCTION

The domain wall (DW) motion induced by a spin-polarized current has been intensively investigated because of its potential applications in emerging magnetic memory[1,2] and logic devices[3,4]. Prototype of such a memory device has been fabricated and demonstrated at the IBM Almaden[1,2] with potential to replace the conventional hard disk memory because of higher storage density and higher speed. The present version of the racetrack memory using NiFe film as the magnetic nanowire in which relatively high current density ($\sim 10^{12}$ A/m$^2$) is required to drive the DW motion. The current density in the order of $10^{12}$ A/m$^2$ may degenerate the performance of device due to Ohmic heating and cause much power consumption[5]. As a result, the reduction of the driving current density is one of the most important issues to commercialize the racetrack memory and take place of the conventional hard disk. Recently, theoretical and preliminarily experimental works have proved that perpendicularly magnetized magnetic thin films revealed a lower current density required to induce DW displacement as compared with in-plane magnetized material, which can reduce Joule heating and protect functional device[6-9]. Some types of the perpendicularly magnetized thin films have been proposed as the promising candidates, such as multilayer films (e.g. Co/Pt[10], Co/Ni[11], CoB/Ni[12]), single layer film (e.g. amorphous TbFeCo film[13,14] ), etc. However, there is still much demand to find proper candidates to fulfill the requirements of the practical devices: 1) reasonable low current density to control the DW motion; 2) high DW velocity for high speed devices; and 3) high structural stability for long-term usage. The amorphous single layer TbFeCo film[13-15] could be considered as an excellent candidate for low current density controlling, but it has not demonstrated a high DW velocity. Besides, it is a fact that amorphous structure of the TbFeCo films is not stable for long-term working under Joule heating. The multilayer films show their advantages on creating the perpendicular magnetic anisotropy (PMA), high DW speed[12], those motivates further studies on them.

It was found that CoFe/Pd multilayers exhibit PMA, controllably low saturation magnetization ($M_s$), and narrow domain wall width ($\Delta$) in our previous report[16]. All these factors are important to reduce the critical current density in spin torque devices. This paper shows an experimental work demonstrating that a low current induced DW motion in a nanowire of CoFe/Pd multilayer with strong perpendicular magnetic anisotropy. Non-adiabatic torque with a high non-adiabatic coefficient is ascribed to be the main contribution to the DWs motion.

## II. EXPERIMENTAL METHOD

A multilayer film with a stack of Ta(5 nm)/Pd(5 nm)/[CoFe(0.4 nm)/Pd(1.2 nm)]$_{15}$/Ta(5 nm) was deposited on thermal oxidized Si(100) substrate. Composition of the CoFe layers was fixed at $Co_{70}Fe_{30}$ (at%). The film was deposited at 1 mTorr Ar pressure by DC magnetron sputtering in an ultrahigh vacuum chamber with base pressure lower than $2.0\times10^{-9}$ Torr. The seed layer of Ta(5 nm)/Pd(5 nm) was applied to promote (111) texture of the magnetic multilayer and Ta(5 nm) was used as a capping layer to avoid oxidation as reported in previous study[16]. The film specimen exhibited a strong perpendicular anisotropy (PMA) with a saturation magnetization $M_s$ = 217 emu/cc, and a uniaxial anisotropy $K_u$ = $2.89\times10^5$ J/m$^3$ [Fig. 1(b)] measured by an alternating gradient force magnetometer (AGFM) under a maximum magnetic field of 2T at room temperature. The value of the $M_s$ was obtained by dividing magnetic moment ($m$) by the total volume of the CoFe/Pd multilayers. The pattern of the nanowire was generated on the substrate by means of electron beam lithography. After the lift-off process, 10μm-long and 600nm-wide nanowires were fabricated. The wire was modified with a triangular contact pad to nucleate the wall, and two Hall crosses with a width of 300 nm for magnetotransport measurements. The anomalous Hall effect (AHE) was used to detect the domain wall motion in nanowires. In the experiment, magnetic field was applied perpendicular to the nanowire plane and the sample was kept in a cryostat at 1.5 K to avoid Joule heating effect on the DW motion. To measure the Hall effect, a DC current of $I_{DC}$=100 µA, which is low enough to guarantee that it does not affect the DW motion, was flowed along the nanowire, and the Hall voltage was measured using nanovolt-meter. Together with the DC current, a pulsed current with a pulse width of 100 µs, and a frequency of 5 kHz was also flowed in the nanowire in order to control the motion of the DWs. A bias-tee was used to set two types of current flowing together in the wire [Fig. 1(a)]. Magnetic domain patterns nucleation were imaged using magnetic force microscopy (MFM) measurement at remanent state.

## III. RESULTS AND DISCUSSION

Fig. 2(a) presents a Hall resistance hysteresis loop ($R_H$-$H$) of the CoFe/Pd multilayer nanowire in absence of the driving pulse current, and in presence of a very small DC current. This illustrates a typical magnetization reversal process under the external magnetic field along the easy axis, similar to that behavior measured in the continuous film sample as shown in Fig. 1(b). A switching field of 0.10 T (1000 Oe) was determined, which is about two times larger than it is in continuous film specimen because of the change in shape anisotropy of the patterned sample. And this reversal is confirmed to be nucleation and movement of the DWs by the MFM imaging in next paragraphs. The switching of the normalized Hall resistance, $R_H$, [point A to point B, Fig. 2(a)], indicates that magnetic DWs start to enter the Hall cross, and the point A is considered as the switching field. Further increasing the field from point B to point C [Fig. 2(a)], a very small change in the $R_H$ is observed, showing that the walls stay pinned in the Hall cross[17]. In this scene, the point B can be defined as pinning field. A quick change of the Hall resistance is again recorded as the field goes beyond the point C, demonstrating the depinning process of the walls from the Hall cross. Based on this evolution, the depinning field (point C), $H_{dep}$, is determined; and two values of depinning field, one in positive field regime and the other one in negative field regime, which are important to extract the driving fields of the magnetization switching in coming discussion, can be found on each $R_H$-$H$ hysteresis loop. Interestingly, the $R_H$-$H$ hysteresis loops are mostly unchanged in absence of the small driven pulse current. When the pulse current is high enough, the change of loops shows the influence of the spin current on the magnetization reversal [Figs. 2(b)], which will be discussed in following paragraphs.

In order to determine the critical current density that is required to drive the motion of DWs, current-switching experiment was conducted as follows: Firstly, the sample was saturated magnetically in a positive field of 1.0 T, then relaxed to the field of -0.075 T to be demagnetized and create the DW. Due to the shape anisotropy, the nucleation pad is easier to reverse and the walls should logically form in the pad prior to appearing in the wire area. Indeed, the MFM imaging shows a stripe-like magnetic domain and Bloch type DWs forming in the nucleation pad (insets of Fig. 3) whilst the straight wire area is still in magnetically uniform state. Subsequently, a pulse current with a density of $J$, and magnitude increased in step is now passed through the wire with intent to inject the walls from the pad to the wire. When the current is high enough (above the critical value, $J_c$), the wall can be driven to move along the wire and through the Hall cross causing a drastic change in the Hall resistance from low to high regime as shown in the Fig. 3. By this experimental, A critical current density of $J_c=1.44\times10^{11}$ A/m$^2$, required to induce the motion of DW in the CoFe/Pd multilayer nanowire with a width of 600 nm, is determined. It is important to notice that this critical current density is significantly lower compared with other common Co-based multilayer nanowires, e.g. Co/Ni (~4-5×10$^{11}$ A.m$^{-2}$)[18-20], Co/Pt (~4×10$^{11}$ A.m$^{-2}$)[21,22], or CoFeB single layer wire (6.2×10$^{11}$ A.m$^{-2}$)[23]. It must be aware of the critical current defined here that prior to completely switching to high-value level, the normalized Hall resistance went through a transitional state at which it only switched to about half value of the high-resistance state. This state seems that the spin current has made it role to drive the wall into the Hall cross, but not high enough to compete the pinning effect of the cross. And we define the critical current density as the value is above which, the spin current is high enough to drive the wall completely through the Hall cross.

Turning back to the $R_H$-$H$ loops [Fig. 2(b)], to avoid crowding, the Fig. 2(b) shows only a few typical curves measured with $J = J_c$, -$J_c$ and $2J_c$. As can be seen from Fig. 2(b), when the driving current density is above the critical value, the loops change drastically with switching field reduced and steps, demonstrating the influence of the spin transfer torque on the DW motion in the nanowire. Furthermore, the $R_H$-$H$ loops at different current polarities also drift from each other due to the different acting direction of the spin torque on the motion of the wall. Note that Joule heating from the current could affect on the switching of the magnetization. However, the measurements were performed in a cryostat remained at 1.5 K, thus, the influence of the Joule heating could be neglected, and the change in the Hall hysteresis loops could be effectively caused by the spin current.

It has been reported that CoFe/Pd multilayers exhibit strong perpendicular magnetic anisotropy (PMA) and narrow domain wall width ($\Delta$)[16], which are considered as the important factors to reduce the driving current density. In addition, the influence of non-adiabatic spin transfer torque ($\beta$) on the DW dynamics is expected to be predominant in narrow DWs due to the higher magnetization gradient[24]. To study the non-adiabatic spin transfer torque effect on CoFe/Pd multilayer nanowires and deduce the quantity of non-adiabatic efficient ($\beta$), we measured the $R_H$-$H$ hysteresis loops with different current densities and current polarities as partially shown in the Fig. 2 (b). From those measurements, the depinning field is obtained as a function of driving pulsed current density in the way described below.

A total of four values of depinning field [$H_{dep}(H^+,I^+)$, $H_{dep}(H^-,I^+)$, $H_{dep}(H^+,I^-)$, $H_{dep}(H^-,I^-)$] were determined from the $R_H$-$H$ loops as shown in the Fig. 2 (b). There are two important effects may be considered as effective fields for determining the $H_{dep}$, namely the spin transfer torque, and the Oersted field. The effects of Oersted field ($H_{Oe}$), spin currents (spin-torque field, $H_{ST}$) have been separated using the formula given in Ref.22.

$$H_{Oe} = \frac{[H_{dep}(H^+,I^+)-H_{dep}(H^+,I^-)]-[H_{dep}(H^-,I^+)-H_{dep}(H^-,I^-)]}{4} \quad (1)$$

$$H_{ST} = \frac{[H_{dep}(H^+,I^+)-H_{dep}(H^+,I^-)]+[H_{dep}(H^-,I^+)-H_{dep}(H^-,I^-)]}{4} \quad (2)$$

The derived $H_{Oe}$ and $H_{ST}$ as a function of the driving current density are shown in Fig. 4. The value of $H_{ST}$ and $H_{Oe}$ corresponds to the mean value of calculated $H_{ST}$ and $H_{Oe}$ over 10 measurements and the error bars demonstrate the standard deviation of the mean value [22]. It can be seen that $H_{ST}$ increases almost linearly as the driving current density increases while the effective Oersted field maintains around zero value. The negative values of effective $H_{Oe}$ shown in Fig. 4 may be caused by the calculated error and the negative values are almost to 0, thus it seems the effective Oersted field makes very little contribution to the current driven domain wall motion. This result confirms the major contribution of the spin torque term to the motion of DWs. From the linear dependence of $H_{ST}$ on the current density, spin-torque efficiency ($\varepsilon$) can be calculated from the slope of the linear dependence ($\epsilon = \mu_0 \Delta H_{ST}/\Delta J$)[22]. As a result, the nonadiabatic efficient $\beta \approx 0.4$ is estimated from the formula $\epsilon = \beta P \hbar \pi / 2 e M_s \Delta$[17]. In comparison with other common Co-based multilayers (e.g. $\beta \approx$ 0.35 in Co/Pt multilayers)[22], the CoFe/Pd multilayers possess relatively higher value of the nonadiabatic coefficient ($\beta \approx 0.4$). The advantages of the studied CoFe/Pd nanowire can be understood by considering the intrinsic properties of the CoFe/Pd multilayer film. It was previously presented that the PMA of the CoFe/Pd multilayer film is essentially contributed by the interfacial magnetic moments of the CoFe/Pd interface due to hybridization between 3d electrons of Fe and 4d electrons of Pd. The CoFe/Pd multilayers with 1.2 nm thickness Pd spacing layers exhibit a low $M_s$ (217 emu/cc), but still remain high anisotropy. A strong $K_u$ is favorable to form narrow Bloch-type walls[16]. Furthermore, Shaw et al. has reported a low Gilbert damping $\alpha$ value in the CoFe/Pd multilayers[25]. A low $M_s$, narrow $\Delta$, low $\alpha$ and high $\beta$ are important parameters to reduce the critical current density for DW motion in nanowire as described by the following equation[7]:

$$j_c = \frac{e\alpha M_s^2 \Delta}{g\mu_B P}\gamma_0 \frac{1}{|\beta-\alpha|} \tag{3}$$

Where, $e$ is the electron charge, $g$ is the gyromagnetic ratio, $\mu_B$ is the Bohr magneton, and $\gamma_0$ is the Landé factor, respectively. It should be noted that the experiments were conducted at low temperature to avoid the effect of Joule heating but it would not change the physical aspects if it is being conducted at room temperature[26].

## IV. CONCLUSIONS

In conclusion, magnetic domain wall motion induced by a low current density of $1.44\times10^{11}$ A/m$^2$ has been observed in the 600nm width CoFe/Pd multilayer nanowire. Depinning experiments showed a reduction of the depinning field and augment of the spin torque term as an effective field with the driving current density. Nonadiabatic spin transfer torque is claimed to be the main contribution to the current induced domain wall motion with a relative high non-adiabatic coefficient up to 0.4. These indicate that CoFe/Pd multilayer has better performances for spin-torque devices comparing with other common Co based multilayers.

**ACKNOWLEDGEMENTS**

We would like to thank the Singapore Agency for Science, Technology and Research (A*STAR) for their financial support to our work through the Grant No. 092 151 0087.

**REFERENCES**

[1] S. S. P. Parkin, M. Hayashi, and L. Thomas, Science **320**, 190 (2008).

[2] M. Hayashi, L. Thomas, R. Moriya, C. Rettner, and S. S. P. Parkin, Science **320**, 209 (2008).

[3] P. Xu, K. Xia, C. Gu, L. Tang, H. Yang, and J. Li, Nature Nanotech. **3**, 97 (2008).

[4] J. Jaworowicz, N. Vernier, J. Ferre, A. Maziewski, D. Stanescu, D. Ravelosona, AS Jacqueline, C. Chappert, B. Rodmacq, and B. Dieny, Nanotechnology **20**, 215401 (2009).

[5] J. H. Ai, B. F. Miao, L. Sun, B. You, An Hu, and H. F. Ding, J. Appl. Phys. **110**, 093913 (2011).

[6] S.-W. Jung, W. Kim, T.-D. Lee, K.-J. Lee, and H.-W. Lee, Appl. Phys. Lett. **92**, 202508 (2008).

[7] H. Szambolics, J.-Ch. Toussaint, A. Marty, I. M. Miron, and L. D. Buda-Prejbeanu, J. Magn. Magn. Mater. **321**, 1912 (2009).

[8] E. Martinez, L. Lopez-Diaz, O. Alejos, and L. Torres, J. Appl. Phys. **106**, 043914 (2009).

[9] S. Fukami, T. Suzuki, N. Ohshima, K. Nagahara, and N. Ishiwata, J. Appl. Phys. **103**, 07E718 (2008).

[10] I. M. Miron, G. Gaudin, S. Auffret, B. Rodmacq, A. Schuhl, S. Pizzini, J. Vogel, and P. Gambardella, Nat. Mater. 9, 230 (2010).

[11] T.Koyama. G. Yamada, H. Tanigawa, S. Kasai, N. Ohshima, S. Fukami, N. Ishiwata, Y. Nakatani, and T. Ono, Appl. Phys. Express 1, 101303 (2008).

[12] D-T. Ngo, N. Watanabe, H. Awano, Jpn. J. Appl. Phys. 51, 093002 (2012).

[13] S. Li, H. Nakamura, T. Kanazawa, X. Liu, and A. Morisako, IEEE Trans. Magn. 46, 1695 (2010).

[14] D-T. Ngo, K. Ikeda and H. Awano, Appl. Phys. Express 4, 093002 (2011).

[15] D. Bang, and H. Awano, Appl. Phys. Express 5, 125201 (2013).

[16] D.-T. Ngo, Z.-L. Meng, T. Tahmasebi, X. Yu, E. Theong, L. H. Yeo, A. Rusydi, G. C. Han and K.-L. Teo, J. Magn. Magn. Mater. **350**, 42 (2014).

[17] O. Boulle, M. Kläui, J. Kimling, U. Rüdiger, P. Warnicke, C. Ulysse, and G. Faini, Phys. Rev. Lett. **101**, 216601 (2008).

[18] H. Tanigawa, T. Koyama, G. Yamada, D. Chiba, S. Kasai, S. Fukami, T. Suzuki, N. Ohshima, N. Ishiwata, Y. Nakatani, and T. Ono, Appl. Phys. Express. **2**, 053002 (2009).

[19] D. Chiba, G. Yamada, T. Koyama, K. Ueda, H. Tanigawa, S. Fukani, T. Suzuki, N. Ohshima, N. Ishwata, Y. Nakatani, and T. Ono, Appl. Phys. Exp. **3**, 073004 (2010).

[20] T. Koyama, D. Chiba, K. Ueda, K. Kondou, H. Tanigawa, S. Fukami, T. Suzuki, N. Ohshima, N. Ishiwata, Y. Nakatani, K. Kobayashi, and T. Ono: Nature Materials. **10**, 194 (2011).

[21] T. A. Moore, I. M. Miron, G. Gaudin, G. Serret, S. Auffret, B. Rodmacq, A. Schuhl, S. Pizzini, J. Vogel, and M. Bonfim: Appl. Phys. Lett. **93**, 262504 (2008).

[22] J. Heinen, O. Boulle, K. Rousseau, G. Malinowski, M. Kl¨aui, H. J. M. Swagten, B. Koopmans, C. Ulysse, and G. Faini: Appl. Phys. Lett. **96**, 202510 (2010).

[23] S. Fukami, T. Suzuki, Y. Nakatani, N. Ishiwata, M. Yamanouchi, S. Ikeda, N. Kasai, and H. Ohno: Appl. Phys. Lett. **98**, 082504 (2011).

[24] G. Tatara, H. Kohno, and J. Shibata, Phys. Rev. Lett. **92**, 086601 (2004).

[25] J. M. Shaw, H. T. Nembach, and T. J. Silva: Phys.Rev. B. **85**, 054412 (2012).

[26] M. Hayashi, Y. Nakatani, S. Fukami, M. Yamanouchi, S. Mitani, and H. Ohno: J. Phys. :Condens. Matter. **24**, 024221 (2012).

**FIGURES**

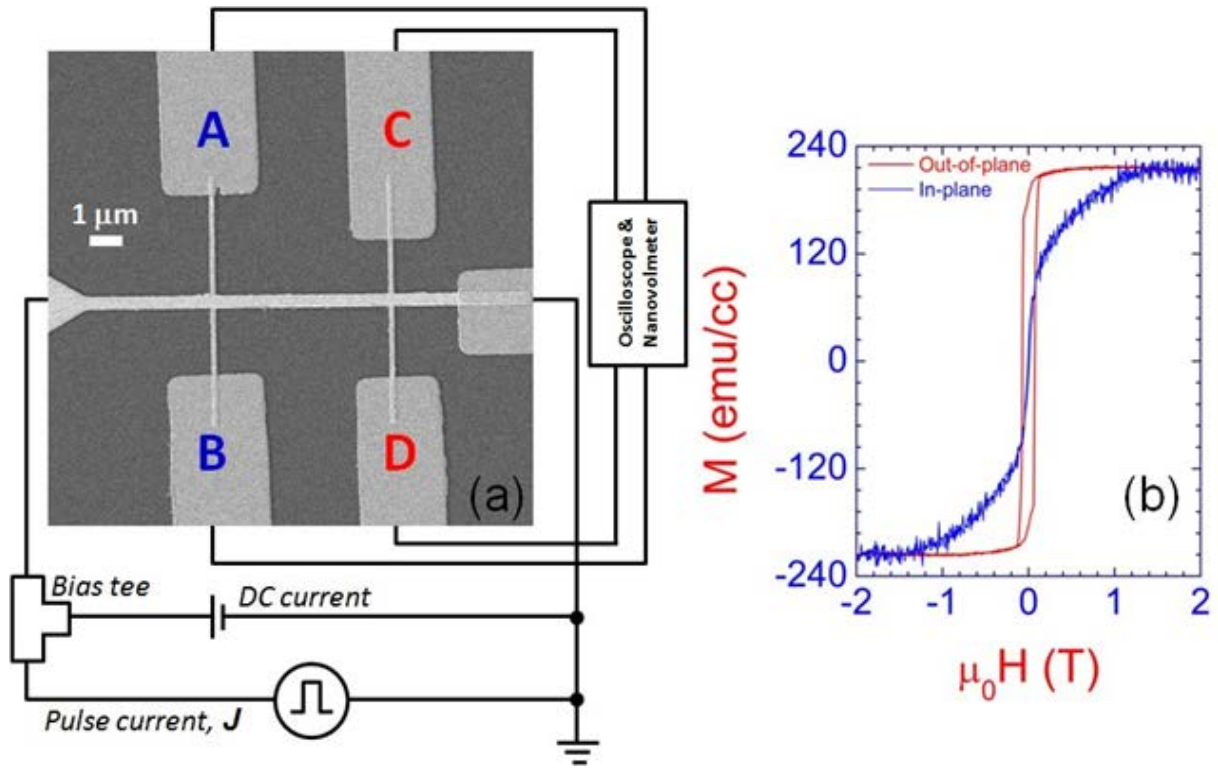

Fig. 1. (a) A SEM micrograph of the nanowire device with a width of 600 nm and the schematic diagram of the experimental setup; A, B, C, D are just used to indicate four electrodes, which are connected to the nano-voltmeter to measure Hall voltage. (b) Hysteresis loops of the continuous film specimen measured by AGFM.

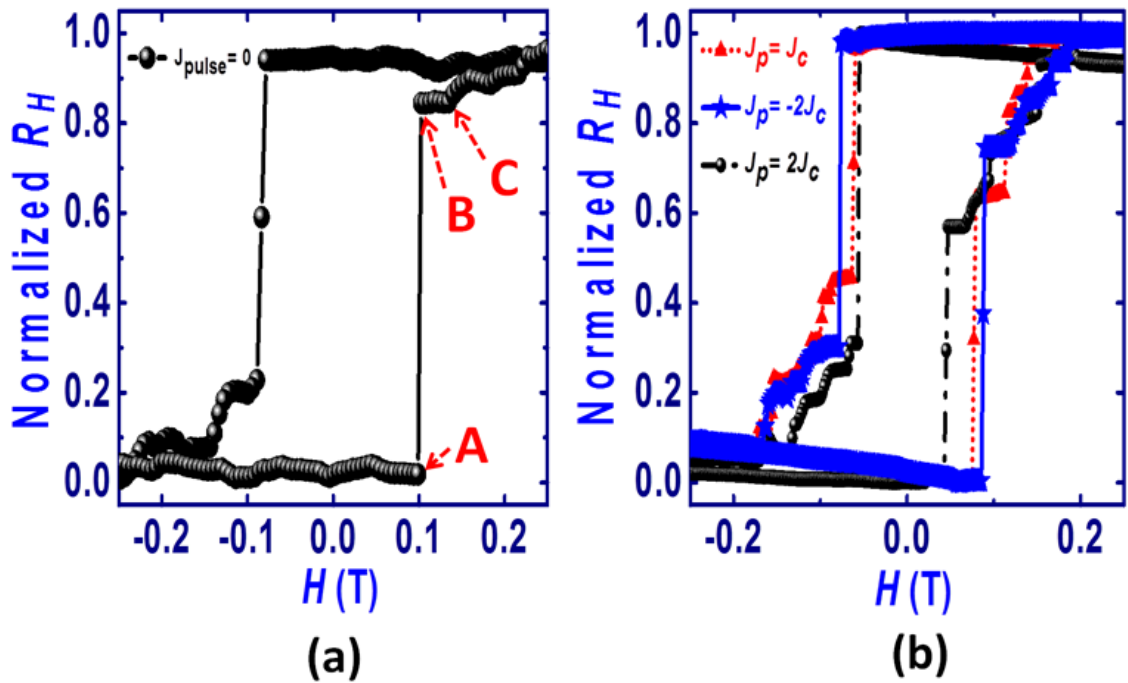

Fig. 2. (a) Normalized Hall resistance hysteresis loops ($R_H$-$H$) measured without the driving pulse current. (b) the $R_H$-$H$ loops obtained at a pulse current density $J=-2J_c$ (solid line), and $J=J_c$ (dot line) and $J=2J_c$ (dash and dot line).

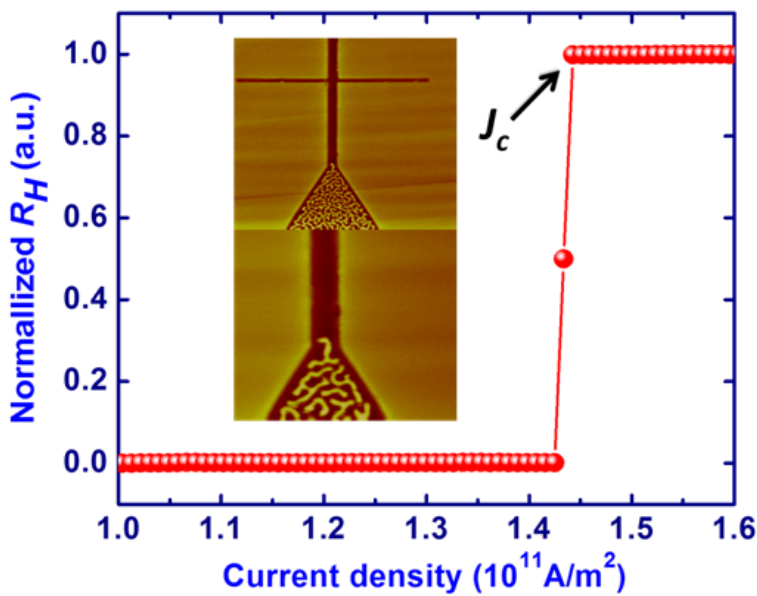

Fig. 3. Current-switching characteristic to determine the critical current density for driving the wall. The inset shows domain walls formed in the nucleation pad at the switching field prior to flowing the driving current.

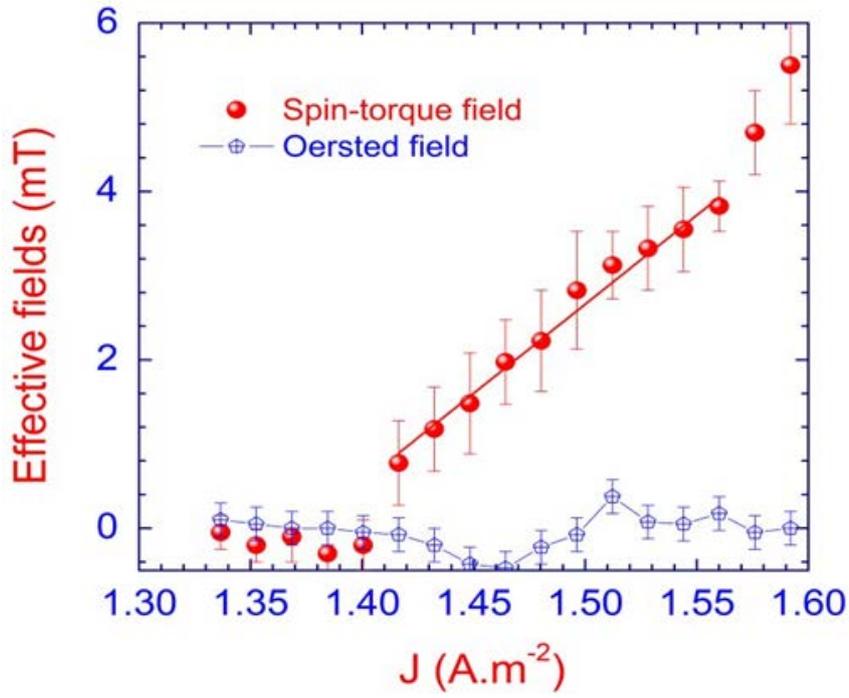

Fig. 4. Spin-torque field ($H_{ST}$) and Oersted field ($H_{Oe}$) as functions of the injected current density. The value of $H_{ST}$ and $H_{Oe}$ corresponds to the mean value of calculated $H_{ST}$ and $H_{Oe}$ over 10 measurements and the error bars demonstrate the standard deviation of the mean value.